# Pacification of thermocapillary destabilization of a liquid film in zero gravity through the use of an isothermal porous substrate

Aneet Dharmavaram Narendranath[1]

[1]Department of mechanical engineering-engineering mechanics, Michigan Technological University, Houghton, MI, USA

**ABSTRACT**

Thin liquid films on isothermal substrates, where film is flat and parallel to the substrate, succumb to thermocapillary instabilities and rupture forming local hot-spots. These instabilities are called long wavelength instabilities and are specific to aspect ratios where the liquid film mean thickness is several orders of magnitude less than the substrate characteristic dimension. In the absence of stabilizing gravitational acceleration, the growth rate of thermocapillary instabilities is further intensified, driving the film to rupture at an even earlier time.

Numerical simulations of zero gravity dynamics of Newtonian liquid films on a solid, horizontal, isothermal substrate are conducted. When the solid, isothermal substrate is replaced with one that possesses one-dimensional porosity, is fully saturated with the same fluid as that of the liquid film and is deep enough to accommodate all of the liquid on it, it is observed that destabilizing spatial modes are damped thereby preventing rupture and the film lifespan is prolonged. This nonlinear evolution is visualized and quantified using recurrence plots.

**ACRONYMS**

| | |
|---|---|
| $h_0$ | Mean film thickness [mm] or [m] |
| $h$ | Film profile, $h = h(x,t)$ [mm] or [m] |
| x | Span-wise coordinate |
| X | Span-wise coordinate (dimensionless) |
| L | Span of substrate [mm] or [m] |
| IS-sub | Isothermal solid substrate |
| IP-sub | Isothermal porous substrate |
| $t$ | Time (dimensional) [s] |
| T | Time (dimensionless) |
| $t_v$ | Viscous time scale, $t_v = {h_0}^2/\upsilon$ [s] |
| $\upsilon$ | Kinematic viscosity of liquid film [m$^2$/s] |
| $\gamma_t$ | Gradient of surface tension ($\sigma$) with respect to temperature, $d\sigma/dT$ [N/m/K] |
| $\Delta T$ | Temperature gradient along interface (dimensional) [K] |
| $T_\infty$ | Ambient temperature (dimensional) [K] |
| $T$ | Temperature (dimensionless), $(T^{*i} - T_\infty)/\Delta T$ |
| $T^{*i}$ | Interfacial temperature (dimensional) [K] |
| $T^i$ | Interfacial temperature (dimensionless) |
| $\mu$ | Dynamic viscosity [N-s/m$^2$] |
| $u_0$ | Characteristic velocity (dimensional) [m/s] |
| $\alpha$ | Thermal diffusivity of liquid film [m$^2$/s] |
| $k_t$ | Thermal conductivity of liquid film [W/m-K] |
| $k_p$ | Permeability parameter as defined in literature (Davis et al. 1999) for Darcy velocity. [mm] or [m] |
| Ca | Capillary number, $\mu u_0 / \sigma$ (dimensionless) |
| Ma | Marangoni number, $\frac{\gamma_t \Delta T}{2 \mu u_0} Ca^{2/3}$ (dimensionless) |
| S | Surface tension number, $Ca^{1/3}/3$ (dimensionless) |
| Bi | Biot number at interface, $\alpha h_0 / k_t$ (dimensionless) |
| N$_{pm}$ | Dimensionless porosity, $\frac{k_p}{h_0} Ca^{-2/3}$ (dimensionless) |

**INTRODUCTION**

Dynamics of and patterning in thin liquid films are central to many important problems in engineering, geophysics, biophysics etc. (Craster et al., 2008). A liquid film on an isothermal substrate is destabilized in zero gravity through the long wave instability of thermocapillary fingering. The lack of gravitational acceleration leads to quicker destabilization than under the influence of terrestrial or non-zero gravitational acceleration (Narendranath et al., 2014). A schematic of a liquid film on an isothermal solid substrate is depicted in figure 1, along with some of the stabilizing and destabilizing effects.

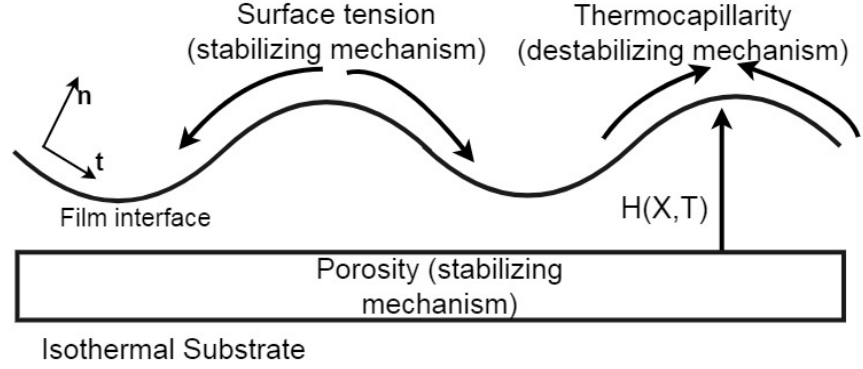

**Figure 1: Various stabilizing and destabilizing mechanisms that affect a liquid film on an isothermal solid substrate, where film is flat and parallel to the substrate. Periodic boundary conditions are used on the left and right edge. In this article, evaporation and concomitant vapor recoil effects are neglected.**

A nonlinear evolution (partial differential) equation derived from the Navier-Stokes equations and the energy conservation equation through a scaling and long wavelength approximation (Burelbach et al., 1988, Ajaev, 2012) is solved numerically to study dynamics of liquid films in zero gravity (zero gravitational acceleration) and a porous substrate induced stabilizing mechanism. This equation allows for the study of the influence of individual mechanisms by engaging required non-linear terms through non-zero dimensionless coefficients (see Acronyms table). The rupture of liquid films (local thinning and exposure of underlying substrate) on isothermal substrates is governed by a balance between destabilizing thermocapillarity and stabilizing surface tension and gravity. In the absence of gravity, long-wave instability driven rupture occurs when thermocapillarity overwhelms surface tension (Narendranath et al., 2014).

Following the model put forth by previously (Davis et al., 1999) or more recently (Liu et al., 2017), the effect of an isothermal porous substrate on zero-gravity evolution of non-evaporating liquid films, without disjoining pressures, is studied. Here, we only consider the case where the substrate is fully saturated with the liquid film and hence effect of wettability, contact angles within the substrate and partial saturation on film dynamics is not considered. Evolution of spatial modes is visualized using recurrence plots and recurrence rate quantification. There exist other accounts in literature that study the effect of spreading of an axisymmetric droplet under the effect of injection or suction of fluid through a slot (Momoniat et al., 2010), the dynamic control of falling films through injection and suction (Thompson et al., 2016) and the spreading of droplets on a deep porous substrate (Liu et al., 2017). These however did not take thermocapillary driven instabilities in the absence of gravitational acceleration, into account. The effect of thermocapillarity, which was excluded from previous studies on films on porous substrates is included in the current study.

**NONLINEAR EVOLUTION EQUATION**

The dimensionless, nonlinear time evolution of a non-evaporating liquid film's interface thickness, $H(X,T)$, in zero gravity is provided by the nonlinear partial differential equation 6. A subscript signifies a partial derivative. The derivation of the evolution equation is summarized below. Inclusion of vital interfacial normal and shear stress balances has been discussed in greater detail in literature (Ajaev, 2012). Starting with the global mass conservation equation:

$$\frac{\partial h}{\partial t} + \frac{\partial Q}{\partial x} = w \quad (1)$$

Where, quantities with dimensions, h, x, t and w are film thickness, h=h(x,t), lateral coordinate, time and percolation Darcy velocity respectively. $Q$ is the interfacial normal and shear stress balance leading to surface tension driven stabilization and thermocapillary led destabilization and fingering.

$$Q = \frac{\sigma}{3\mu}\frac{\partial^2 h}{\partial x^2} - \frac{\gamma_t}{2\mu}h^2\frac{\partial T^{*i}}{\partial x} \quad (2)$$

The following scaling arguments are used to non-dimensionalize the governing equations of fluid dynamics in the lubrication approximation limit of small aspect ratio, $h_0/L \ll 1$.

$$x \sim X\, h_0 Ca^{-1/3} \quad (3)$$

$$t \sim T\, t_v \quad (4)$$

Scaling the film thickness as $h \sim H\, h_0$ with leading order dimensionless interfacial temperature (Oron et al., 1997) given as $T^i = 1/(1 + Bi\, H)$

$$\frac{\partial T^{*i}}{\partial x} = \frac{\partial}{\partial x}(\Delta T T^i + T_\infty) = \Delta T \frac{\partial T^i}{\partial x} = \Delta T \frac{\partial}{\partial x}\left(\frac{1}{1+Bi\, H}\right) \quad (5)$$

In the derivation of the evolution equation (equation 6), gravity driven stabilization is neglected. Davis and Hocking use a surface tension time scale in their work (Davis et al. 1999). However, a viscous time scale is used to develop the evolution equation (Burelbach et al., 1988; Narendranath et al. 2014) to capture thermocapillary effects. Choice of a time scale based on a general characteristic velocity instead of a viscous scale yields the same evolution equation.

$$\partial_T H + \left[MaBi\frac{H^2}{(1+BiH)^2}H_X + SH^3 H_{XXX}\right]_X = N_{pm}SH_{XX} \quad (6)$$

In equation 6, The term, $\partial_T H$, tracks the time-evolution of the film interface. Thermocapillary interfacial instabilities are captured by the nonlinear term, $\left[MaBi\frac{H^2}{(1+BiH)^2}\right]_X$. Stabilizing effect of surface tension is captured by $[SH^3 H_{XXX}]_X$. The capillary effect of a porous substrate, as modeled by Darcy flow equation, is captured by the term $N_{pm}SH_{XX}$. The non-dimensional quantities Ma, Bi are the Marangoni and interface Biot number. The non-dimensional surface tension number is S, the one-dimensional porosity of the substrate is captured in the non-dimensional number $N_{pm}$. These non-dimensional numbers (included in the acronyms table) are set to: Ma=5.00, Bi=1.00, S=100.0 which are values routinely used in literature (Oron et al., 2000). It is the magnitude of these values relative to each other, which determines film interface evolution. We have simulated the dynamics of dichloromethane films previously (Narendranath et al., 2014) with a Ma-S ratio ($m$) of 0.1092. The current value of ($m$) of 0.05 leads to a slower thermocapillary evolution as compared to $m$=0.1092.

Equation 6 is solved with periodic boundary conditions, with a period of $2\pi/k$ and a small perturbation initial condition, viz., $1 - 0.1cos(kx)$. Here, $k$ is the fastest growing wavenumber (reciprocal of wavelength) shown in equation 7 is determined from a linear stability analysis.

$$k = \left[\frac{1}{S}\left\{\frac{Ma}{(1+Bi)^2} - N_{pm}\right\}\right]^{0.5} \quad (7)$$

The value of $N_{pm}$ is chosen to be 0.001 to ensure very little divergence from the fastest growing wavelength for a porous substrate with respect to a non-porous substrate. This value of $\frac{N_{pm}}{S} = 10^{-5}$ ensures that initial film dynamics are strongly affected by the surface tension term as compared to capillarity due to the porous substrate. In subsequent section, $\frac{N_{pm}}{S} = 10^{-5}$ is termed "weak porosity".

The numerical solution is obtained using Wolfram *Mathematica* and its method of lines option along with the LSODA solver (Momoniat et al., 2010) and is validated with results in literature (Narendranath et al., 2014).

**RESULTS**

**Dynamics of film evolution in zero gravity**

Thermocapillary instabilities affecting a liquid film on an isothermal solid substrate **(IS-sub)** in zero gravity are shown in figure 2. The film profile approaches rupture due to thermocapillarity at approximately T=1650. In our simulations, rupture is considered to have been achieved when the local film thickness is ≈ 1000 times smaller (Narendranath et al., 2014) than the initial mean film thickness. Rupture of liquid films is caused by the development of higher frequencies that manifest as new thermocapillary spatial modes that evolve from the initial condition. When an isothermal porous substrate **(IP-sub)** was used in the simulations, the film persisted without rupturing until T=3500.0 (figure 3) at which point, the simulation was halted.

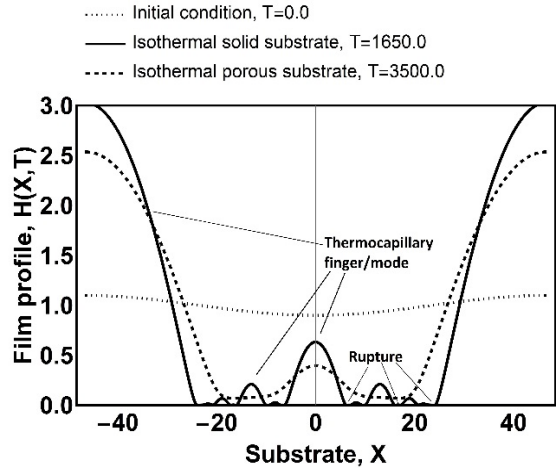

**Figure 2: Evolution of a liquid film towards thermocapillary driven rupture from a cosine initial condition. An isothermal porous substrate allows for the film to persist without rupture for longer than with an isothermal solid substrate.**

**Visualizing spatial modes through recurrence plots**

To visualize the similarity in film dynamics and related spatial modes at different time instances with different substrates, a recurrence matrix is calculated and plotted. The recurrence matrix (Marwan et al., 2007), which is a sparse matrix, is calculated using equation 8.

$$R_p = \theta\big(\epsilon - \|q_j - q_i\|\big) \qquad (8)$$

Where ε is a properly chosen cut-off distance, $q_j, q_i$ are states of the dynamical system at closely spaced, immediate spatial locations $i,j$ and $\theta$ is the unit step function. The $\|q_j - q_i\|$ is the Euclidean distance (or L-2 Norm) between states $q_j, q_i$. When the recurrence matrix is plotted as an array or matrix plot, a recurrence plot (RP) is available. This RP reveals all the times when the phase space trajectory of the dynamical system visits roughly (with proximity $\epsilon$) the same position in the phase space. When a state recurs (in this case, when consecutive points on the film profile are within the cut-off distance of each other), the recurrence matrix is populated with "1". Plotting this sparse matrix shows the locations of spatial recurrence (darkened positions correspond to "1") and the presence of multiple modes through distortions in the diagonal or off-diagonal lines. Recurrence plots are prominently used in analyzing stochastic data such as machine tool vibrations, medical scans and financial data and are an alternative to phase portraits. A phase portrait is not particularly useful to study film evolution because different points in the film evolve at different rates and to different amplitudes. The recurrence plot allows for a normalization, visualization and analysis of spatially evolving modes at different time instances.

The evolution of a liquid film on the IS-sub in zero gravity, is shown in figure 4. A cut-off distance ε of 0.0005 signifies that the relative (spatial) difference of 0.05% between a current state and the previous state is treated as a recurrence or a recurrent state. Physically, this cut-off distance corresponds to ≈ $1\mu m$ or less. This distance is nearly equal to rupture thickness in our simulations.

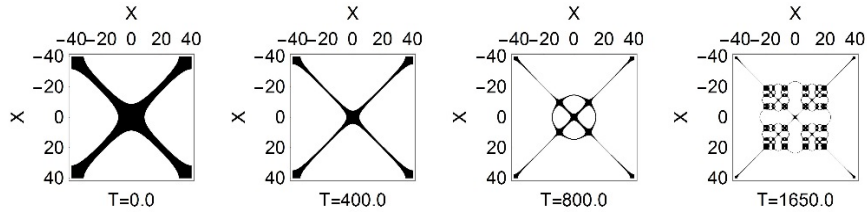

**Figure 4: Recurrence plot for a liquid film evolving on an ISS in zero gravity. Cut-off length, ε=0.0005.**

In contrast to figure 4, the recurrence plot for a liquid film on the IP-sub in zero gravity is depicted in figure 5. Higher spatial modes are damped when a porous substrate is used. This is seen as a lack of gaps in the diagonals and off-diagonal segments of the matrix plot. These gaps exist when an IP-sub is used (fig 4), which in this case means that there is a continuing development of destabilizing thermocapillary modes.

A difference in dynamics between liquid films on the two different substrates is visualized through a cross-recurrence plot (equation 9) in figure 6. In equation 9, $p, q$ are states of two different dynamical systems, i.e, film profile on a solid substrate and that on a porous substrate, that are being compared for similarity, at the same time. The cut-off distance used in figure 6 is ε=0.0005. In other words, this cross recurrence plot captures the nearness of film dynamics of a liquid film on IS-sub to the liquid film on IP-sub, at different spatial location, evolving in time, within 0.05% of each other

$$CR_p = \theta(\epsilon - \|p - q\|) \qquad (9)$$

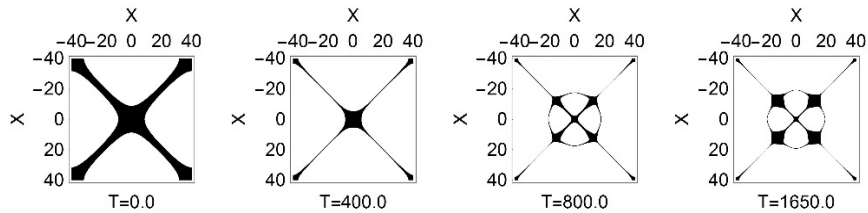

**Figure 5: Recurrence plot for a liquid film evolving on an IP-sub in zero gravity. Cut-off length, ε=0.0005.**

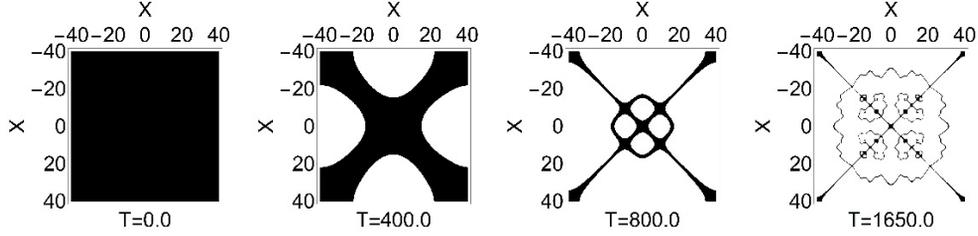

**Figure 6: Cross recurrence plots (CRP) of liquid film states for cut-off length, ε=0.0005.  This is a visual comparison of the film dynamics on the IS-sub vs that on the IP-sub for a recurrence of states within 0.05% of each other.  Since with the IS-sub, the film ruptures at T=1650.0, CRP comparisons are made only until T=1650.0**

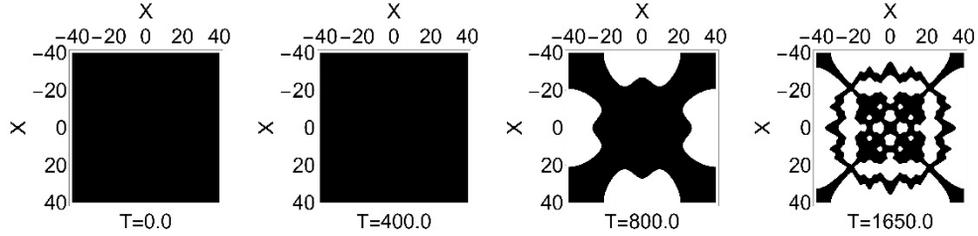

**Figure 7: CRP of liquid film states for cut-off length, ε=0.05.  This is a visual comparison of the film dynamics on the IS-sub vs that on the IP-sub for a recurrence of states within 5% of each other.**

A larger value of the cut-off distance, ε=0.05, is used and the closeness in film dynamics of a film on the IS-sub vs the film on the IP-sub is depicted in figure 7. In other words, figure 7 captures the recurrence of spatio-temporal states within 5% of each other.  To quantify this closeness in dynamics, a rate of recurrence (RR%), which is one of the many recurrence quantification parameters (Marwan et al. 2007) is calculated and tabulated in table 1.   The rate of recurrence is simply the percentage of '1' in the sparse recurrence matrix.  A greater percentage of '1' suggests a greater rate or percentage of recurrence.

**DISCUSSION**

In this short communication, numerical simulations are conducted for a Newtonian liquid film on an isothermal solid (IS-sub) and isothermal porous substrate (IP-sub), in zero gravity. Here, the film is flat and parallel to the substrate. The porous substrate is treated as having one-dimensional depth that is fully saturated with the same fluid as that of the liquid film and is deep enough to accommodate all of the liquid on it that is sufficient to store the entire volume of the liquid film that rests on it. A nonlinear evolution equation derived using a long wave expansion is solved as an initial value problem with periodic boundary conditions.  The liquid film is under the influence of stabilizing surface tension and destabilizing thermocapillarity, with zero gravitational acceleration.  With the IS-sub, the film ruptures through the creation and growth of thermocapillary modes and concomitant local thinning.

Next a weakly porous substrate is chosen to ensure that the fastest growing wavelength stays nearly the same as that for the non-porous substrate.  It is observed that this has the effect of damping of spatial thermocapillary modes and prolonging film lifespan without rupture. Rupture of the film, on the isothermal porous substrate, is not observed in the time-frame simulated.

Recurrence plots are used to analyze the temporal evolution of spatial thermocapillary modes, with and without a porous substrate. An isothermal porous substrate leads to the damping of higher spatial modes. Cross-recurrence plots (CRPs) are used to visualize a difference in states between a liquid film on an isothermal solid substrate and an isothermal porous substrate. A recurrence rate (RR%) measuring the nearness of dynamics of the film states with different substrates reveals that with a weakly porous substrate, the film dynamics although similar at early stages of evolution, lead to different states at later stages.

Future work will include studies of zero gravity liquid film/coating pattern fidelity through the use of recurrence plots and recurrence quantification. Windowing of recurrence plots will be used to quantify temporal evolution of spatial points of interest.

| Liquid film on ISS vs IPS with $\varepsilon=0.05$ | | | Liquid film on ISS vs IPS with $\varepsilon=0.0005$ | | |
|---|---|---|---|---|---|
| Time (non-dimensional) | Recurrence Rate (RR%) | Significance | Time (non-dimensional) | Recurrence Rate (RR%) | Significance |
| 0.0 | 100.0 | Similar dynamics | 0.0 | 100.0 | Similar dynamics |
| 400.0 | 100.0 | Similar dynamics | 400.0 | 33.66 | Dynamics within 33.66% of each other |
| 800.0 | 45.94 | Dynamics within 45.94% of each other | 800.0 | 6.86 | Dynamics within 6.86% of each other |
| 1650.0 (time rupture of film on ISS) | 39.79 | Dynamics within 39.79% of each other | 1650.0 | 5.29 | Dynamics within 5.29% of each other |

**Table 1: Closeness of dynamics, calculated through a rate of recurrence, of a liquid film on ISS vs the liquid film on IPS. Two different cut-off distances, are used to describe this closeness.**

**ACKNOWLEDGEMENTS**

The author wishes to acknowledge the support provided by the department of mechanical engineering-engineering mechanics at Michigan technological university. The author also thanks the reviewers and the journal editorial team for providing excellent comments to help improve this paper, in timely fashion.